\documentclass[conference]{IEEEtran}
\IEEEoverridecommandlockouts
\usepackage{amsmath,amssymb,amsfonts}
\usepackage{algorithmic}
\usepackage{graphicx}
\usepackage{textcomp}
\usepackage{xcolor}
\def\BibTeX{{\rm B\kern-.05em{\sc i\kern-.025em b}\kern-.08em
    T\kern-.1667em\lower.7ex\hbox{E}\kern-.125emX}}
\usepackage[backend=biber,style=ieee]{biblatex}
\begin{document}

\title{Chained Prompting for Better Systematic Review Search Strategies}

\author{\IEEEauthorblockN{Fatima Nasser}
\IEEEauthorblockA{\textit{Electrical and Computer Engineering} \\
\textit{American University of Beirut}\\
Beirut, Lebanon \\
fhn05@mail.aub.edu}
\and
\IEEEauthorblockN{Fouad Trad}
\IEEEauthorblockA{\textit{Electrical and Computer Engineering} \\
\textit{American University of Beirut}\\
Beirut, Lebanon \\
fat10@mail.aub.edu}
\and
\IEEEauthorblockN{Ammar Mohanna}
\IEEEauthorblockA{\textit{Electrical and Computer Engineering} \\
\textit{American University of Beirut}\\
Beirut, Lebanon \\
am288@aub.edu.lb}
\and
\IEEEauthorblockN{Ghada El-Hajj Fuleihan}
\IEEEauthorblockA{\textit{Department of Internal Medicine} \\
\textit{American University of Beirut}\\
Beirut, Lebanon \\
gf01@aub.edu.lb}
\and
\IEEEauthorblockN{Ali Chehab}
\IEEEauthorblockA{\textit{Electrical and Computer Engineering} \\
\textit{American University of Beirut}\\
Beirut, Lebanon \\
chehab@aub.edu.lb}
}

\maketitle

\begin{abstract}
Systematic reviews require the use of rigorously designed search strategies to ensure both comprehensive retrieval and minimization of bias. Conventional manual approaches, although methodologically systematic, are resource-intensive and susceptible to subjectivity, whereas heuristic and automated techniques frequently under-perform in recall unless supplemented by extensive expert input.
We introduce a Large Language Model (LLM)-based chained prompt engineering framework for the automated development of search strategies in systematic reviews. The framework replicates the procedural structure of manual search design while leveraging LLMs to decompose review objectives, extract and formalize PICO elements, generate conceptual representations, expand terminologies, and synthesize Boolean queries. In addition to query construction, the framework exhibits superior performance in generating well-structured PICO elements relative to existing methods, thereby strengthening the foundation for high-recall search strategies.
Evaluation on a subset of the LEADSInstruct dataset demonstrates that the framework attains a 0.9 average recall. These results significantly exceed the performance of existing approaches. Error analysis further highlights the critical role of precise objective specification and terminological alignment in optimizing retrieval effectiveness.
These findings confirm the capacity of LLM-based pipelines to yield transparent, reproducible, and high-performing search strategies, and highlight their potential as scalable instruments for supporting evidence synthesis and evidence-based practice.
\end{abstract}


\section{Introduction}

Evidence-based research represents the standard approach adopted by healthcare professionals to guide patient care, therapeutic decisions, and clinical practice \parencite{cumpston2019updated}. A central component of this research paradigm is the literature search (LS), conducted across relevant databases, which provides the foundational basis for producing high-quality systematic reviews (SRs) \parencite{aromataris2014systematic}.  
In the context of systematic reviews, the literature search (LS), involves the formulation of a search strategy derived from the review question \parencite{eriksen2018impact}. The review question describes the objective of the study and, consequently, LS determines the body of evidence available for subsequent analysis. Because the quality and scope of the SR are contingent upon the data retrieved, the LS constitutes a critical methodological component. It must therefore be designed with rigor to ensure robustness and to minimize bias when selecting the relevant articles \parencite{rethlefsen2021prismas}. 

In recent years, LS has become increasingly systematic, necessitating the use of professional search strategies (SS), which are frequently developed through librarian-mediated search services \parencite{bramer2017optimal}. A common approach to search strategy development in SRs is to translate the research question into structured concepts, identify relevant keywords and subject headings for each concept, and combine them with Boolean and proximity operators to create a comprehensive, reproducible search strategy across databases. However, the design of such strategies is often constrained by time, as information specialists may lack deep subject-matter expertise and must rapidly acquire sufficient familiarity with diverse domains while still producing effective search strategies. Traditionally, this process relies on a conceptual, or concept-based, subjective approach, which presents several limitations: (1) uncertainty as to when a strategy can be considered “complete”; (2) increased complexity and higher risk of errors as the number of queries grows; (3) excessive retrieval requiring subsequent narrowing; and (4) high demands on time and effort \parencite{hausner2012routine}. Employing more objective methodologies for constructing search strategies offers a potential means of overcoming these challenges.

Harnessing the capabilities of LLMs has the potential to significantly enhance the effectiveness of LS. By integrating LLMs capabilities, these systems can deliver precise results, leading to better overall performance \parencite{tian2023opportunities}. Language models (LMs) are built to understand and generate human language by considering contextual information within word sequences. When applied to query writing, the use of LLMs generally falls into multiple approaches \parencite{zhu2024largelanguagemodelsinformation}. One of these commonly used approaches is prompting. In prompting, the techniques entail the formulation of targeted instructions that guide the responses of LLMs, thereby providing both adaptability and interpret-ability \parencite{zhu2024largelanguagemodelsinformation}. 

In this study, we present and evaluate a prompt engineering pipeline designed to generate search strategies that are efficient, transparent, and reproducible. This is done by:

\begin{itemize}
    \item  Introducing a chained prompt engineering pipeline based on GPT-4o-mini that automates SR's search strategy construction, including PICO extraction, concept identification, keyword generation, and Boolean query construction. 

    \item Achieving a state-of-the-art performance on the LEADSInstruct dataset \parencite{wang2025foundation}, significantly outperforming existing methods, while generating more accurate and structured PICO elements. 

    \item Providing error analysis that highlights key challenges for low recall results, and demonstrating that improved objectives can fully recover recall. 
\end{itemize}

\section{Related Work}

Bibliographic databases are highly structured with sophisticated indexing. This requires a more strategic search approach compared to a simple Web search \parencite{chandler2019cochrane}. In this section, we present the evolution of constructing a search strategy (SS) and the efforts to automate the process.

\subsection{Manual Approach}
When constructing an SS, particularly for clinically oriented questions, librarians commonly employ the PICO framework (Population, Intervention, Comparison, Outcome) to decompose the research objective into its essential components. This framework facilitates the identification of core concepts that inform keyword selection. To achieve comprehensive retrieval, selected terms are typically expanded to include synonyms, lexical variants, and semantically related expressions, thereby mitigating the risk of omitting relevant studies. In addition, many bibliographic databases, including PubMed, Medline, and Embase, utilize controlled vocabularies and subject headings (e.g., MeSH terms) to standardize indexing and enhance retrieval precision. Accordingly, an effective search strategy integrates both free-text keywords and controlled vocabulary terms, combined through Boolean operators (OR, AND, NOT) to logically structure and refine the search \parencite{chandler2019cochrane}.

\subsection{Heuristic Approach}
litsearchr \parencite{grames2019automated} is a tool that provides a quasi-automated workflow for the construction of search strategies. The process begins with the formulation of a concise, concept-driven Boolean query (e.g., derived from the PICO framework), followed by aggregation of search results across multiple databases and removal of duplicates. Candidate terms are then extracted using the Rapid Automatic Keyword Extraction (RAKE) algorithm from titles, abstracts, and author- or database-supplied keywords. A co-occurrence network is subsequently constructed to rank terms by node strength, with genetic-algorithm heuristics applied to determine thresholds for retaining influential terms. Researchers review and refine these suggestions by incorporating additional variants (e.g., stems and near synonyms), after which litsearchr automatically generates optimized Boolean queries that incorporate stemming, redundancy reduction, and multilingual support \parencite{grames2019automated}.

\subsection{LLMs Approach}
LEADS  \parencite{wang2025foundation} is a fine-tuned LLM for systematic reviews that can be used to automate and optimize the construction of search queries. Through prompt engineering applied to the titles and abstracts of included studies, research questions are reformulated within the PICO framework. Based on these generated PICO structures, GPT-4o is prompted to extract candidate population and intervention terms, which are then systematically combined to produce high-recall Boolean queries. An extension of this approach, \textbf{LEADS+ensemble}, enhances coverage and robustness by generating multiple queries (ten per review), executing them independently, and aggregating the results without duplicates \parencite{wang2025foundation}. As such, for the \(n\)-th study, they apply a population term set \(P_n=\{p^n_1,\dots,p^n_M\}\) and an intervention term set \(I_n=\{i^n_1,\dots,i^n_M\}\), where \(M\le 10\). Within a study, extracted terms are combined using the AND boolean operator \(\land\), while across studies they are aggregated using the OR boolean operator \(\lor\). This yields the population and intervention sub-queries in equations \eqref{eq:SP} and \eqref{eq:SI}. Moreover, the final synthetic query is given in equation \eqref{eq:S}.

\begin{align}
S_P &= S^1_P \lor S^2_P \lor \cdots \lor S^N_P, \label{eq:SP}\\
S_I &= S^1_I \lor S^2_I \lor \cdots \lor S^N_I, \label{eq:SI}\\
S   &= S_P \land S_I. \label{eq:S}
\end{align}

Here \(S^n_P = p^n_1 \land p^n_2 \land \cdots \land p^n_M\) and \(S^n_I = i^n_1 \land i^n_2 \land \cdots \land i^n_M\).

In contrast, our proposed workflow follows the structure of the traditional manual approach to systematic review searching, but leverages the capabilities of LLMs through chained prompt engineering. The developed pipeline produces a well-formulated search strategy directly from a clearly defined objective, achieving substantially higher recall without requiring additional seed studies or extensive human intervention. In contrast to existing methods, our system demonstrates that even without ensembling, a streamlined pipeline based on GPT-4o-mini achieves superior recall, highlighting the efficiency and scalability of the approach.

\section{Methodology}
\begin{figure}[!b]
    \centering
    \includegraphics[width=\linewidth]{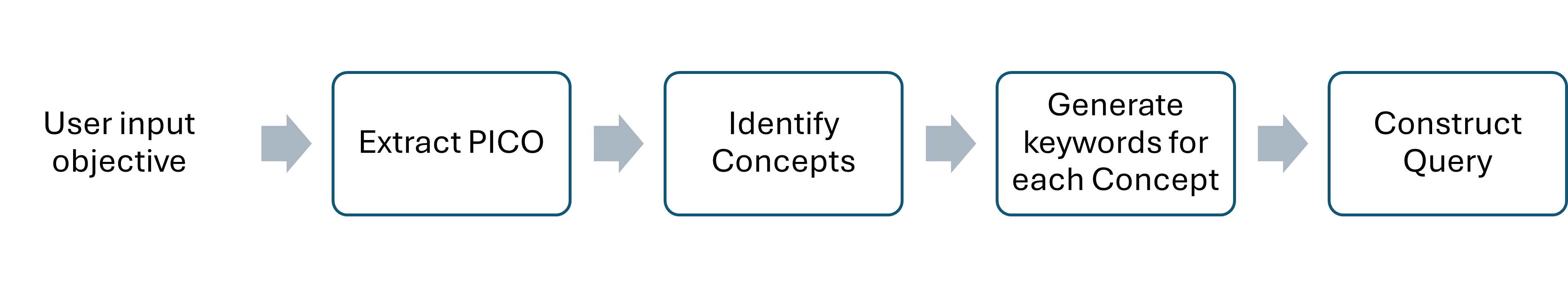}
    \caption{Overview of the proposed chain prompting model for systematic review search strategy generation}
    \label{fig:chain_prompting}
\end{figure}

When building an SS, precision is essential, and the task should be divided into multiple stages to achieve the best outcome. In this context, rather than relying on a single prompt to generate a search strategy from a given objective, a multi-step prompting approach is more effective \parencite{wu2022ai}. This method produces a richer and more focused set of keywords and terms, whereas a single-prompt approach often yields broader and less precise results. In our proposed chain prompting model Fig.~\ref{fig:chain_prompting}, the process is decomposed into sub-tasks, with the LLM focusing on a narrower and well-defined objective at each stage.

At a conceptual level, the proposed pipeline parallels the traditional manual approach, proceeding through four primary stages that define its structure. The process begins with the specification of a clearly articulated objective, encompassing both the research question and the focus of the SR. From this objective, PICO elements are extracted as the initial step. These elements are subsequently mapped onto a set of concepts, for which the model generates candidate keywords. In the final stage, the accumulated elements are integrated into a comprehensive Boolean query, which is executed across databases to retrieve relevant citations. The following section provides a detailed description of the model and elaborates on the design of each component within the pipeline.

\subsection{Model Architecture}

The pipeline was implemented using GPT-4o-mini (OpenAI) as the underlying LLM for all stages. Each step was guided by a combination of a system prompt, which defined the model’s role and constraints, and a user prompt, which specified the task requirements. The outputs were designed to be structured and hierarchically organized, ensuring consistency across stages. This structured format not only facilitated downstream processing but also allowed the model to capture complex and extended contexts more effectively by leveraging precise attention mechanisms within an organized framework \parencite{liu2024enhancing}.

\subsection{Search Strategy Prompts Chain}

\subsubsection{Step 1: Extract PICO Elements}

\begin{figure*}[!h]
    \centering
    \includegraphics[width=0.95\textwidth]{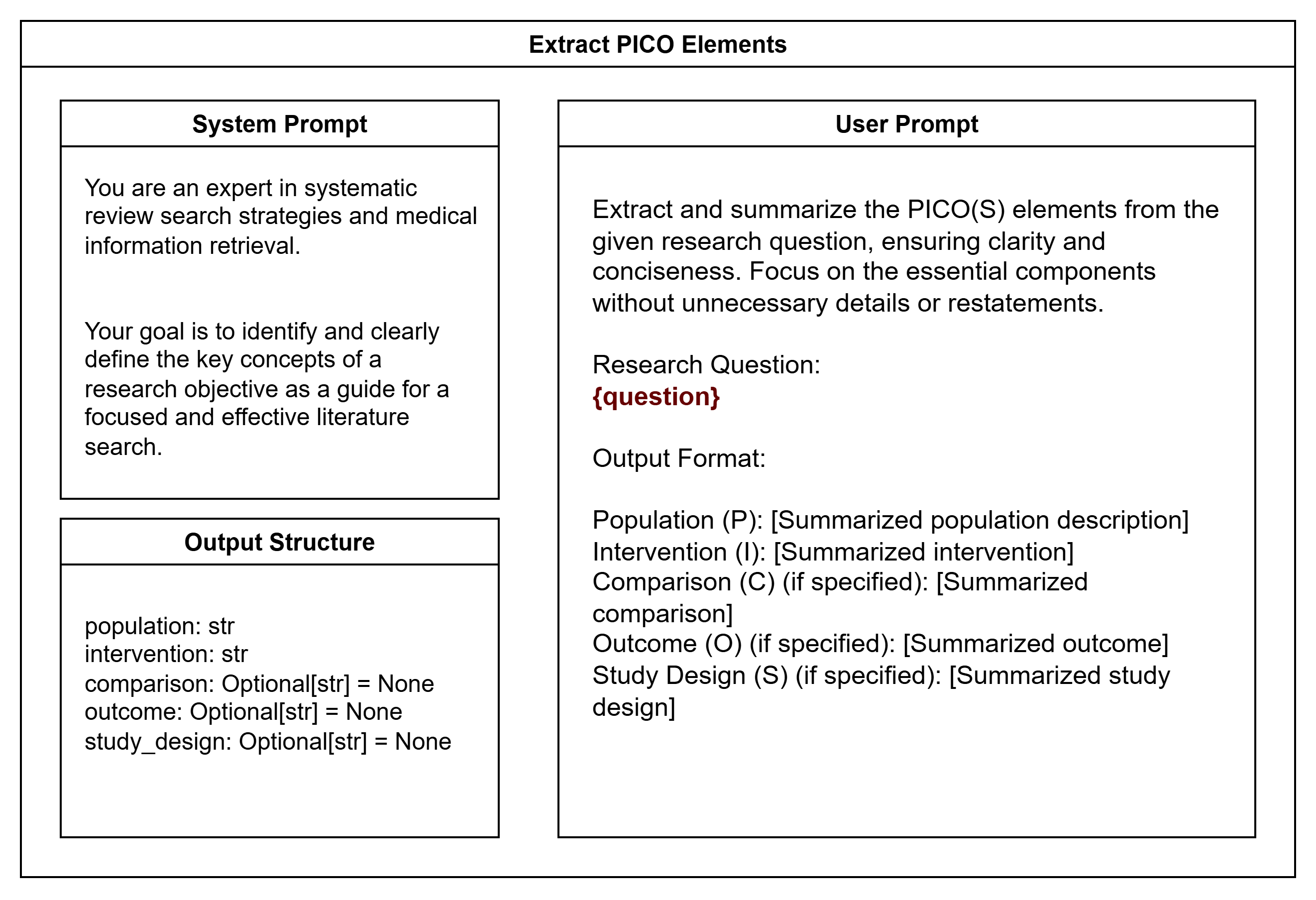}
    \caption{PICO elements prompting diagram}
    \label{fig:pico_prompt}
\end{figure*}

In the first step, the model is prompted to decompose the research question/objective into structured PICO elements (Population, Intervention, Comparison, Outcome). This ensures a clear and systematic representation of the clinical focus, which serves as the foundation for subsequent concept identification as shown in Fig.~\ref{fig:pico_prompt}.

\subsubsection{Step 2: Identify Concepts}

\begin{figure*}[!h]
    \centering
    \includegraphics[width=0.95\textwidth]{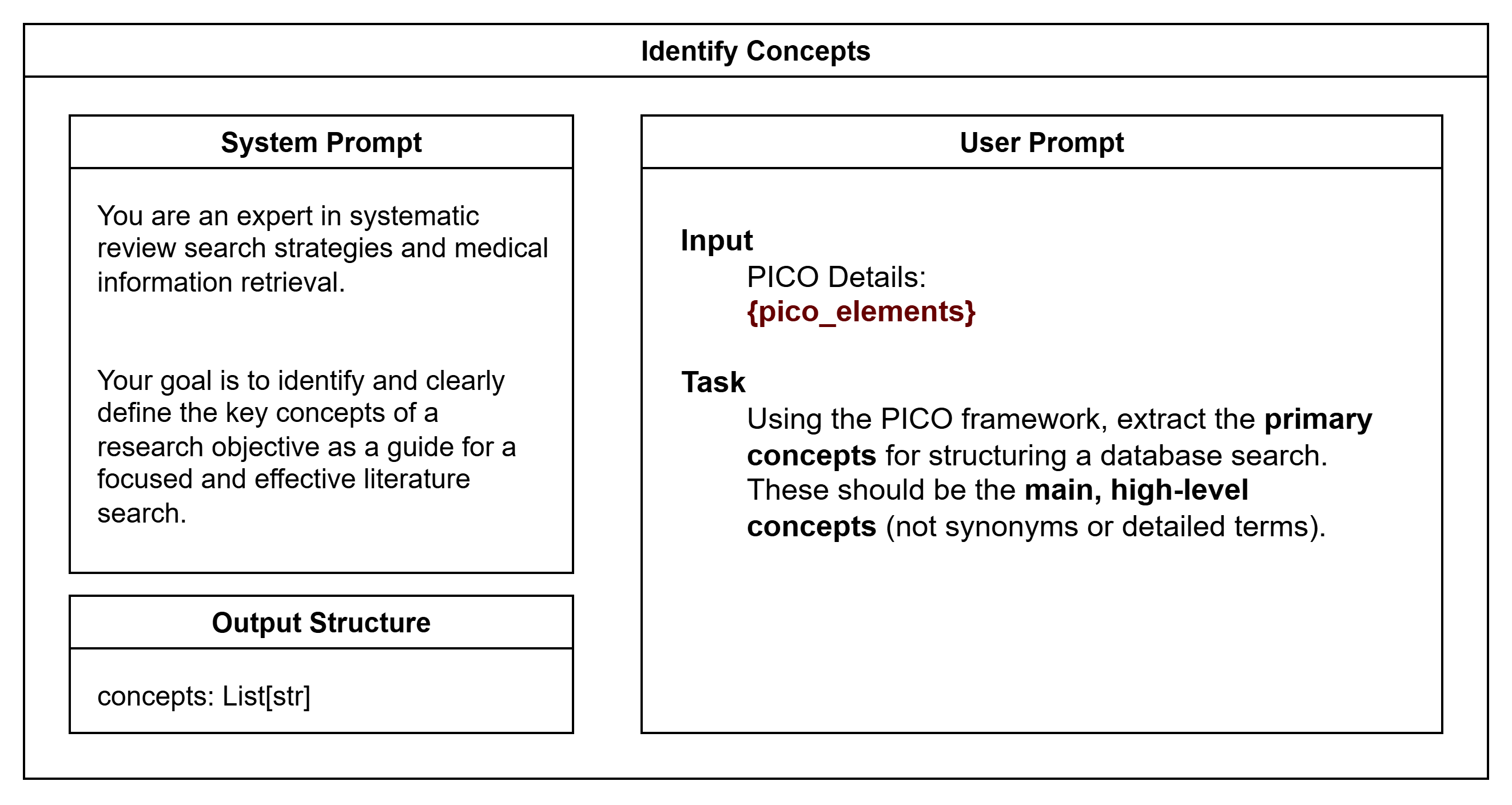}
    \caption{Concepts prompting diagram}
    \label{fig:concepts_prompt}
\end{figure*}

At this stage, the extracted PICO components are expanded into domain-specific concepts. The prompts direct the model to refine broad elements into more precise conceptual categories, clarifying the informational scope and identifying the search terms to be incorporated as shown in Fig.~\ref{fig:concepts_prompt}.

\subsubsection{Step 3: Generate Concepts Keywords}

\begin{figure*}[!h]
    \centering
    \includegraphics[width=0.95\textwidth]{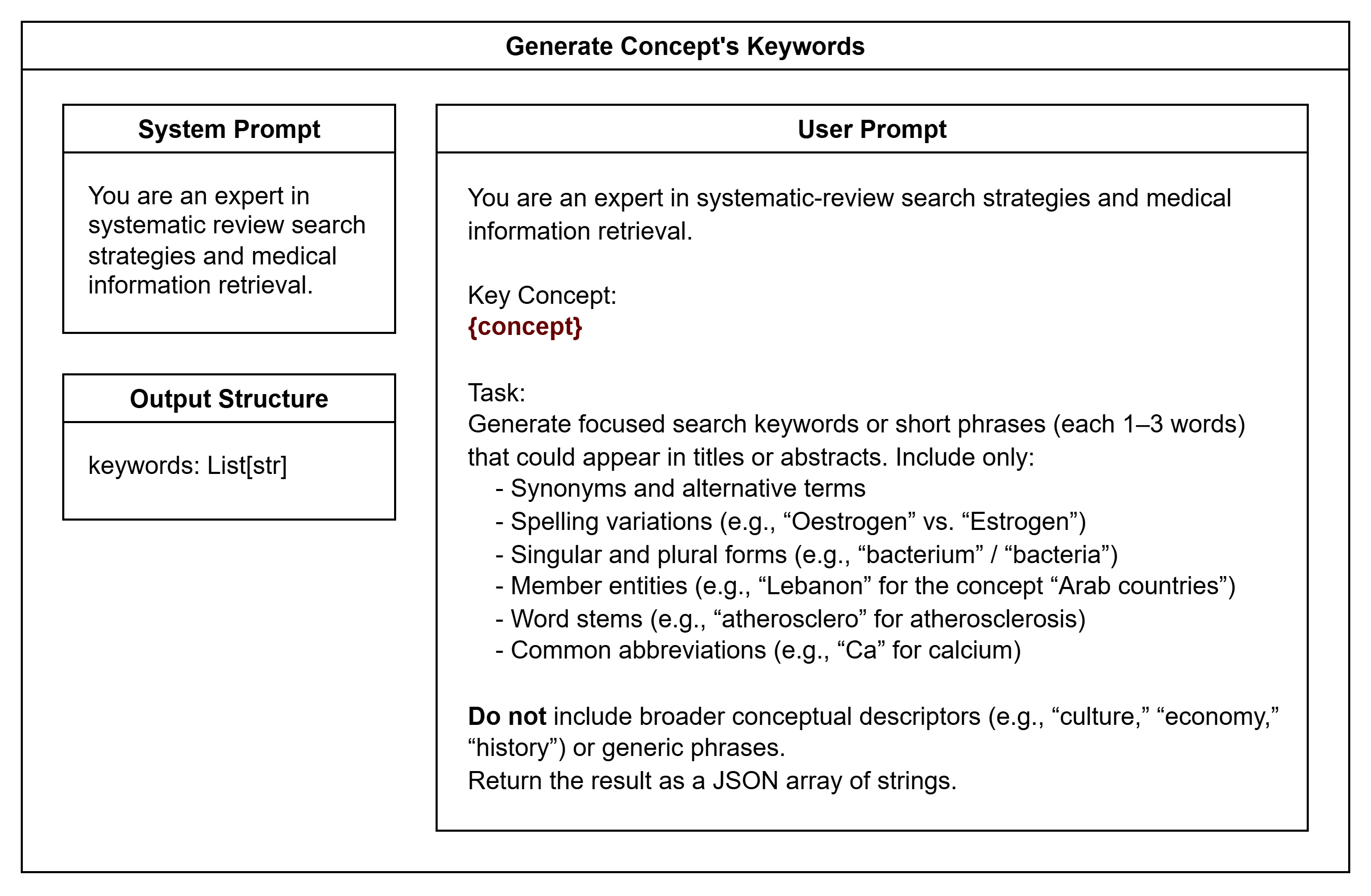}
    \caption{Keywords prompting diagram}
    \label{fig:keywords_prompt}
\end{figure*}

For each concept, the model generates a set of relevant keywords and variations. This step enriches the search space by including synonyms, spelling variants, and related terminologies, ensuring higher recall in database queries as shown in Fig.~\ref{fig:keywords_prompt}.

\subsubsection{Step 4: Construct Query}

\begin{figure*}[!h]
    \centering
    \includegraphics[width=0.95\textwidth]{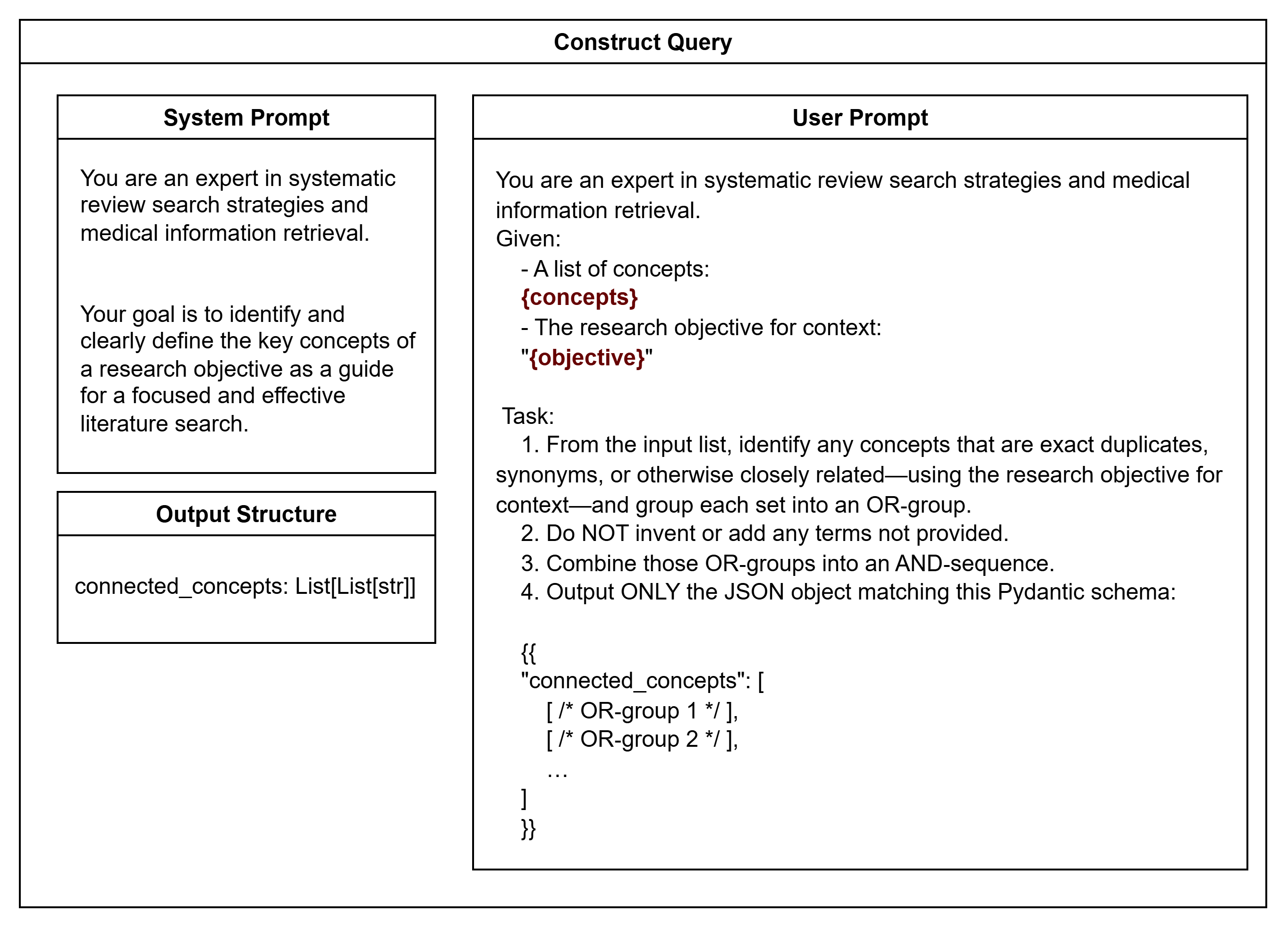}
    \caption{Query construction prompting diagram}
    \label{fig:query_prompt}
\end{figure*}

In the final stage, the previously identified concepts are integrated, in alignment with the original research objective, through prompts that instruct the model to construct Boolean expressions and apply proximity syntax consistent with established systematic review guidelines as shown in Fig.~\ref{fig:query_prompt}. Each concept is expanded into its associated keywords, which are subsequently combined using the OR operator, before being linked across concepts through logical connectors to form the complete search strategy.

\section{Evaluation and Analysis}
\subsection{Dataset}
To evaluate the proposed pipeline, we employed the LEADSInstruct study search publication dataset \parencite{wang2025foundation}. This dataset consists of PubMed-indexed SRs, each accompanied by the review title and abstract, PICO elements generated through prompt engineering by the LEADSInstruct authors, and the PubMed identifiers (PMIDs) of the studies included in the review. To ensure that the evaluation was restricted to high-quality SRs, we applied a series of inclusion criteria, yielding a final subset of 81 reviews:

\begin{enumerate}
    \item Excluded SRs whose search strategies returned more than 1,000 citations, due to PubMed per one API call limits.
    \item Retained only SRs published between 2012 and 2016, to ensure methodological quality, reliance on PICO-based search strategies, and absence of AI involvement in their development.
    \item Removed SRs that returned API errors during retrieval.
\end{enumerate}

\subsection{Results}
\begin{figure}[htbp]
    \centering
    \includegraphics[width=\linewidth]{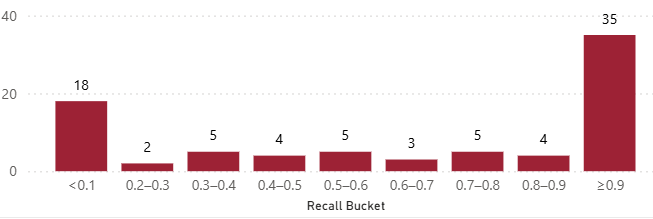}
    \caption{Recall distribution for pipeline from LEADs PICO}
    \label{fig:leads_pico_recall}
\end{figure}

Because the dataset included PICO elements generated via prompt engineering but did not provide explicit review objectives, we initiated evaluation from Step 2 of the pipeline, in which concepts are derived from the supplied PICO elements. As summarized in Table~\ref{tab:recall_comparison}, this approach yielded an average recall of 0.62 across 81 systematic reviews. When search strategies with very low recall ($<$0.2) were excluded, the average recall increased to 0.80. The recall distribution further indicated that 35 search strategies achieved perfect recall (all target articles retrieved), whereas 18 strategies resulted in a recall of 0 (no target articles retrieved), as shown in Fig.~\ref{fig:leads_pico_recall}.  
 
\begin{figure}[htbp]
    \centering
    \includegraphics[width=\linewidth]{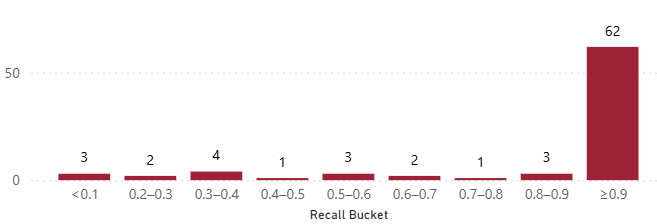}
    \caption{Recall distribution for Original pipeline}
    \label{fig:full_pipeline_recall}
\end{figure}

To improve performance, we applied our \textbf{full pipeline}, while adding an additional step of reformulating a structured objective from each title and abstract before extracting PICO elements. Using our designed prompt to extract PICO elements significantly enhanced the results, raising the average recall to 0.87, and to 0.90 when strategies with recall below 0.2 were excluded as shown in Table~\ref{tab:recall_comparison}. The distribution of results also improved remarkably, with 62 out of 81 search strategies achieving perfect recall, and only 3 remaining with a recall of 0, as shown in Figure~\ref{fig:full_pipeline_recall}.

\subsection{Evaluation and Comparison}

In this study, we focused our evaluation primarily on \textbf{recall} because, at the search strategy stage of a systematic review, the principal objective is to maximize retrieval of all potentially relevant citations rather than to minimize noise. Since our experiments were conducted on already completed SRs, the key benchmark for assessing model performance was whether the final included articles were successfully captured. Precision and screening burden, while important, are downstream concerns that are addressed within the broader SR pipeline we are building. In our previous work \parencite{Trad2025Streamlining}\parencite{nasser2024accelerating}, we have developed and automated both \textbf{title/abstract screening} and \textbf{full-text screening} modules, which directly handle the refinement of the retrieved set. Thus, recall at the search stage ensures completeness, while subsequent pipeline components ensure efficiency and relevance.

As shown in Table~\ref{tab:recall_comparison}, our model substantially outperforms existing approaches in terms of recall. The baseline GPT-4o achieves only 0.10 average recall, highlighting the limitations of direct prompting. LEADS improves this to 0.24, while \textbf{LEADS+ensemble}, reported only on search strategies with recall greater than 0.2, achieves an average recall of 0.82 by aggregating multiple queries. To ensure a fair comparison with these methods, we also report our results under the same condition (excluding cases with recall $<$0.2). Under this setting, our pipeline starting from the PICO elements provided by LEADS achieves a recall of 0.80, while our \textbf{full pipeline}, which includes the objective reformulation step, reaches 0.90.

When considering all strategies (without filtering), our method achieves an average recall of 0.87 for the full pipeline. This corresponds to a \textbf{362\% relative improvement} over LEADS (0.24 $\rightarrow$ 0.87) and an \textbf{8\% gain} over LEADS+ensemble under comparable conditions (0.82 $\rightarrow$ 0.90).

Notably, both LEADS and our approach rely on prompt engineering to generate PICO elements. We compared two prompts: the original LEADS prompt and our own engineered version. Our prompt design produces substantially better PICO extraction, as reflected in the higher recall achieved using our PICO (0.87) compared to LEADS’ PICO (0.62).
\begin{table}[!t]
\centering
\caption{Comparison of average recall across different models}
\label{tab:recall_comparison}
\resizebox{\linewidth}{!}{%
\begin{tabular}{lcc}
\hline
\textbf{Model / Approach} & \textbf{Average Recall} & \textbf{Average Recall ($>$0.2 only)} \\
\hline
GPT-4o & 0.10 & -- \\
LEADS & 0.24 & -- \\
LEADS+ensemble & -- & 0.82 \\
Our model (LEADS' PICO start) & 0.62 & 0.80 \\
Our model (Full pipeline) & 0.87 & 0.90 \\
\hline
\end{tabular}%
}
\end{table}

These results demonstrate that our method not only surpasses ensemble-based strategies but also underscores the importance of prompt design. By directly comparing LEADS’ PICO prompt with our engineered prompt, we show that improved PICO formulation yields higher recall (0.87 vs. 0.62). Effective prompt engineering is therefore critical for accurately capturing review objectives and achieving robust, high-coverage retrieval in systematic reviews.
\subsection{Error Analysis}
To investigate the limitations of the pipeline, we performed an error analysis on a random sample of five search strategies with recall values below 0.5. This analysis identified the following issues:

\begin{enumerate}
    \item \textbf{Terminology mismatch:} In one case, the article did not use standard medical terminology (e.g., ``workers'' instead of ``adults''), which limited the effectiveness of keyword extraction and query matching.
    \item \textbf{Dataset quality issue:} One of the selected articles was not a systematic review and therefore should not have been included in the evaluation dataset, highlighting noise in the benchmark.
    \item \textbf{Objective formulation errors:} In three cases, the automatically derived objectives from the title and abstract were insufficient to capture the true scope of the review. When we manually replaced these with objectives derived from the corresponding systematic review protocols and re-applied the pipeline, all required articles were successfully retrieved, raising the recall to 1.0 in each case.
\end{enumerate}

The analysis indicates that the primary sources of error arise from (i) inconsistent terminology within the literature, (ii) limitations in the quality of the evaluation dataset, and (iii) shortcomings in objective reformulation when relying exclusively on titles and abstracts. Notably, the results demonstrate that providing more accurate and structured objectives (e.g., those derived from systematic review protocols) markedly improves performance. This finding underscores the importance of providing a well-formulated objective as input to the pipeline in professional use cases. If we remove these cases then we would be left with a recall of around 90\%.

\section{Conclusion}
In this study, we introduced a chained prompt engineering pipeline for the development of search strategies in systematic reviews, integrating the methodological rigor of manual approaches with the efficiency of large language models. By decomposing the task into sequential stages, PICO extraction, concept identification, keyword expansion, and query construction, the pipeline establishes a transparent and reproducible workflow. Evaluation on the LEADSInstruct dataset demonstrated substantial performance gains, achieving a relative improvement of 362\% over LEADS (from 24\% to 87\%), and a 10\% improvement over LEADS+ensemble (from 80\% to 90\%) under comparable conditions. Beyond recall, the pipeline consistently produced more accurate and structured PICO elements than the baseline methods, ensuring that the downstream concept expansion and query formulation are built on a stronger foundation. The error analysis emphasized the importance of precise objective formulation and robust terminology handling for optimizing performance. These findings highlight the potential of LLM-driven approaches to improve the quality and scalability of literature searches, reducing dependence on manual expertise and streamlining the systematic review process. Future work will focus on extending the approach to additional bibliographic databases beyond PubMed and refining strategies for query reformulation.

\printbibliography

\end{document}